\documentclass[11pt,reqno]{amsart}
\usepackage{amsfonts,amsmath,amssymb}
\newtheorem{thm}{Theorem}[section]

\usepackage{amssymb}
\usepackage{color}
\usepackage{pstricks}
\usepackage{pst-node}

\title[Calogero--Moser operators in infinite dimension]{Calogero--Moser operators in infinite dimension }
\author{ A.N. Sergeev}\address{Department of Mathematical Sciences,
Loughborough University, Loughborough LE11 3TU, UK}
\email{A.N.Sergeev@lboro.ac.uk}

\author{A.P. Veselov}
\address{Department of Mathematical Sciences,
Loughborough University, Loughborough LE11 3TU, UK  and Moscow State University, Moscow 119899, Russia}
\email{A.P.Veselov@lboro.ac.uk}

\begin{document}

\maketitle

\begin{abstract}
Various infinite-dimensional versions of the Calogero--Moser operator are discussed. The related class of Jack--Laurent symmetric functions is studied. In the special case when parameter $k=-1$ the analogue of Jacobi--Trudy formula is given and the relation with representation theory of Lie superlagebra $\mathfrak{gl}(m,n)$ is discussed.
\end{abstract}

%\tableofcontents 

\section{Introduction} 

Calogero--Moser problem describes the interacting particles on the line with the inverse square potential, or in the Sutherland case $\sinh^{-2}$-potential.
In the quantum trigonometric case we have the following Calogero--Moser--Sutherland (CMS) operator
$$L_k^{(N)} = -\sum_{i=1}^N
\frac{\partial^2}{\partial
x_{i}^2}.+\sum_{i<j}^N \frac{2k(k+1)}{\sinh
^2(x_{i}-x_{j})}.$$ 
 It has an eigenfunction $$\Psi_0= \prod_{i<j}^N \sinh^{-k} (x_i-x_j)$$ with the eigenvalue 
$\lambda_0= k^2N(N-1)/4.$
Its gauged version $\Psi_0^{-1} (L_N-\lambda_0) \Psi_0$ in the exponential coordinates 
$z_i = e^{2x_i}$ has the form
\begin{equation}
\label{CM}
 {\mathcal L}_{k}^{(N)}=\sum_{i=1}^N
\left(z_{i}\frac{\partial}{\partial
z_{i}}\right)^2-k\sum_{ i < j}^N
\frac{z_{i}+z_{j}}{z_{i}-z_{j}}\left(
z_{i}\frac{\partial}{\partial z_{i}}-
z_{j}\frac{\partial}{\partial
z_{j}}\right).
\end{equation}
One can slightly modify it by adding a multiple of the momentum integral
\begin{equation}
\label{CMmod}
\tilde {\mathcal L}_{k}^{(N)}=\sum_{i=1}^N
\left(z_{i}\frac{\partial}{\partial
z_{i}}\right)^2-k\sum_{1\le i < j\le N}
\frac{z_{i}+z_{j}}{z_{i}-z_{j}}\left(
z_{i}\frac{\partial}{\partial z_{i}}-
z_{j}\frac{\partial}{\partial
z_{j}}\right)+k(N-1)\sum_{i=1}^N
z_{i}\frac{\partial}{\partial z_{i}}$$ $$ = \sum_{i=1}^N
\left(z_{i}\frac{\partial}{\partial z_{i}}\right)^2-2k\sum_{i=1}^N\left(\sum_{
j \ne i} \frac{z_{i}z_{j}}{z_{i}-z_{j}}\right) \frac{\partial}{\partial
z_{i}}.
\end{equation}
An important property of this modified operator is its {\it stability}
in the following sense.

Let $P_{N}={\mathbb C}[z_{1},\dots, z_{N}]$ be the polynomial
algebra in $N$ independent variables and $$\Lambda_{N} = \mathbb C[z_{1},\dots, z_{N}]^{S_N}\subset
P_{N}$$ be the subalgebra of symmetric polynomials. For any $M>N$ we have the homomorphisms
$$\phi_{M,N}: \Lambda_M \rightarrow \Lambda_N,$$ sending $z_i$ with $i>N$ to zero.
Then the following diagram is commutative
$$
\begin{array}{ccc}
\Lambda_{N}&\stackrel{\tilde{\mathcal
L}_{k}^{(N)}}{\longrightarrow}&\Lambda_{N} \\ \downarrow
\lefteqn{\varphi_{N,M}}& &\downarrow \lefteqn{\varphi_{N,M}}\\
\Lambda_{M}&\stackrel{\tilde{\mathcal
L}_{k}^{(M)}}{\longrightarrow}&\Lambda_{M} \\
\end{array}
$$ (see \cite {Ma}, example 3 on the page 326). 
This allows us to define an infinite-dimensional version of the CMS operator ${\mathcal L}_{k}^{(\infty)}$ on the space of symmetric functions as follows. 
Consider the inverse limit of $\Lambda_N$ in the category of graded algebras
$$\Lambda=\lim_{\longleftarrow} \Lambda_{N}.$$
By definition, $f \in \Lambda^r$ corresponds to an infinite sequence of elements $f_N \in \Lambda^r_N, \, N=1,2,\dots$ of degree $r$ such that
$\phi_{M,N} f_M = f_N.$
The elements of $\Lambda$
are called {\it symmetric functions}. The power sums
$$p_l = z_1^l + z_2^l + \dots,\,\, l=1,2, \dots$$ is a convenient set of free generators of this algebra, the set
$$p_{\lambda}=p_{\lambda_1}p_{\lambda_2}\dots$$ for all partitions $\lambda$ forms a linear basis in $\Lambda.$
Consider a natural homomorphism $\varphi_{N}:\Lambda \longrightarrow \Lambda_{N}$
 defined by
\begin{equation}
\label{phin}
\varphi_{N}(p_{l})=\sum_{i=1}^N z^{l}_{i}.
\end{equation} 
One can show that there exists  a unique second order differential operator $\tilde {\mathcal L}^{(k)}: \Lambda \to \Lambda$ such that for all $N=1,2,\dots$  the following diagram is commutative
$$
\begin{array}{ccc}
\Lambda&\stackrel{\tilde{\mathcal
L}_{k}^{(\infty)}}{\longrightarrow}&\Lambda\\ \downarrow
\lefteqn{\varphi_{N}}& &\downarrow \lefteqn{\varphi_{N}}\\
\Lambda_{N}&\stackrel{\tilde{\mathcal
L}_{k}^{(N)}}{\longrightarrow}&\Lambda_{N} \\
\end{array}
$$
It has 
 the following explicit form in power sums $p_{a}$ (see \cite{Stanley, Awata}):
 \begin{equation}\label{infinity}
\tilde{ \mathcal L}_{k}^{(\infty)}=\sum_{a,b>0}p_{a+b}\partial_{a}\partial_{b}-k\sum_{a,b>0}p_{a}p_b \partial_{a+b}
 +\sum_{a>0}(a+ak-k)p_a\partial_a,
\end{equation}
where $$\partial_a = a\frac{\partial}{\partial p_a}.$$
 The proof is straightforward with the use of the following identities: 
 $$
 \sum_{i < j}^N(z_{i}+z_{j})\frac{z_{i}^a-z_{j}^a}{z_{i}-z_{j}}=-ap_{a}+\sum_{i=0}^{a-1}p_{i}p_{a-i},
 $$
  $$
 \sum_{i < j}^N\frac{z_{i}^a-z_{j}^a}{z_{i}-z_{j}}=-\frac12ap_{a-1}+\frac12\sum_{i=0}^{a-1}p_{i}p_{a-i-1},
 $$
where $p_0=N.$ 
This is a standard way (going back to Stanley \cite{Stanley}) to lift the (trigonometric) CMS operator to infinite dimension.

We are going to show now that there is another way to do this, which does not require the stability of the operator and works also in the rational and $BC$-case. 

The main idea is that in infinite dimension we have not just one operator but a {\it family depending on the additional parameter} $p_0$ replacing the dimension $N.$ This idea is not new. It was probably first used in $BC$ case by Rains \cite{Rains} (for Koornwinder polynomials) and later by the authors \cite{SV3} to define $BC_{\infty}$ CMS operator, but in $A$-case it was not exploited (probably because of the possibility to stabilize the operators described above).

In the Laurent polynomial case there is no stability and the CMS operators as well as 
the corresponding eigenfunctions (which we call {\it Jack--Laurent symmetric functions}) must depend on the additional parameter.
This observation goes back to Sogo \cite{Sogo1,Sogo2},  but in the finite-dimensional situation he dealt with this dependence is not essential, which is however not the case in infinite dimension.
We discuss the properties of the Jack--Laurent symmetric functions $P_{\lambda,\mu}(k,p_0)$ labelled by two partitions and write down the Pieri formula for them. 

In the last section we discuss the special case $k=-1.$ We show that the limit of Jack--Laurent symmetric functions $P_{\lambda,\mu}(k,p_0)$ when $k\rightarrow -1$ for generic $p_0$ does not depend on  $p_0$ and can be given by the corresponding analogue of Jacobi--Trudy formula. We explain the relation of this with the results by Moens and Van der Jeugt \cite{MVDJ} about characters of the irreducible representations of Lie superalgebra $\mathfrak{gl}(m,n).$

\section{Calogero--Moser operators in infinite dimension}

Let us come back to the original CMS operator (\ref{CM}). Although it is not stable we still can lift it to $\Lambda$ as follows.

Note that in the finite dimensional case 
$$p_0=1+1+\dots+1=N$$ is just the dimension, but in infinite dimension it does not make sense.
The idea is to use $p_0$ as an {\bf additional parameter} and to allow the operators to depend on it.

 \begin{thm}\label{main} 
 {\em i)} There exists  a unique second order differential operator $\mathcal L_{k,p_0}^{(\infty)}: \Lambda \to \Lambda$ polynomially dependent on the additional parameter $p_0$, such that for all $N=1,2,\dots$ and $p_0=N$   the following diagram is commutative
$$
\begin{array}{ccc}
\Lambda&\stackrel{{\mathcal
L}_{k,p_0}^{(\infty)}}{\longrightarrow}&\Lambda\\ \downarrow
\lefteqn{\varphi_{N}}& &\downarrow \lefteqn{\varphi_{N}}\\
\Lambda_{N}&\stackrel{{\mathcal
L}_k^{(N)}}{\longrightarrow}&\Lambda_{N} \\
\end{array}
$$
where  $\mathcal L_k^{(N)}$ is the CMS operator (\ref{CM}).

  {\em ii)} Operator $\mathcal L_{k,p_0}^{(\infty)}$ has the following explicit form in the power sums coordinates $p_{a}:$
  \begin{equation}\label{inf1}
{ \mathcal L}_{k,p_0}^{(\infty)}=\sum_{a,b>0}p_{a+b}\partial_{a}\partial_{b}-k\sum_{a,b>0}p_{a}p_b \partial_{a+b}- kp_0 \sum_{a>0} p_{a} \partial_{a} +(1+k)\sum_{a>0}a p_a\partial_a,
\end{equation}
where as before $\partial_a = a\frac{\partial}{\partial p_a}.$

{\em iii)} Operator $\mathcal L_{k,p_0}^{(\infty)}$ has the following symmetry (duality)
 \begin{equation}\label{sym}
\theta^{-1} \circ \mathcal L_{k,p_0}^{(\infty)} \circ \theta = k \mathcal L_{k^{-1}, k^{-1}p_0}^{(\infty)},
 \end{equation}
where
\begin{equation}\label{theta}
\theta: p_a \rightarrow k  p_a, k  \rightarrow k^{-1}.
 \end{equation}
 \end{thm}
 
Note that we have also a remarkable symmetry between the first and the second terms, so that if  following \cite{Awata} we define $\tilde \partial_a=-\frac{a}{k} \frac{\partial}{\partial p_a}$ the corresponding Hamiltonian will be invariant under swapping $p_a$ and $\tilde \partial_a$ (Fourier duality).
 
The duality $\theta$ changes the parameter $p_0$, which means that it does not work in the finite dimensional situation, when the dimension is fixed. This fact was known already to Stanley and Macdonald, who probably were the first to discover this duality (see \cite{Stanley, Ma}).
We would like to mention that duality (\ref{sym}) looks more elegant than for the stable version (\ref{infinity}), when the momentum operator has to be involved.

The possibility of stabilization is clear from the formula (\ref{inf1}) since parameter $p_0$ appears only as a coefficient at the momentum operator 
$$P= \sum z_i \frac{\partial}{\partial z_i} = \sum p_a \partial_a,$$ 
which is just another quantum integral of the system.

This however is not the case for the rational version of the CMS operator:
\begin{equation}
\label{CMratio}
 {\mathbf L}_{k}^{(N)}=\sum_{i=1}^N
\frac{\partial^2}{\partial
z_{i}^2}-2k\sum_{i < j}^{N}
\frac{1}{z_{i}-z_{j}}\left(
\frac{\partial}{\partial z_{i}}-
\frac{\partial}{\partial
z_{j}}\right),
\end{equation}
for which the corresponding infinite-dimensional version is (in the same notations)
 \begin{equation}\label{inf1ratio}
{ \mathbf L}_{k,p_0}^{(\infty)}=\sum_{a,b\geq 1}p_{a+b-2}\partial_{a}\partial_{b}-k\sum_{a,b\geq 0}p_{a}p_b \partial_{a+b+2} +(1+k)\sum_{a \geq 2}(a-1) p_{a-2}\partial_a.
\end{equation}
Its $p_0$-dependent part is
$$p_0 \partial_1^2 +[(1+k)p_0 - k p_0^2] \partial_2 - 2kp_0 \sum_{b>0}p_b \partial_{b+2},$$
while the momentum operator here has the form
$${\mathbf P}= \sum_{a>0}p_{a-1} \partial_a.$$ 
The problem with infinite dimension can be seen already when we apply the Laplacian
$\Delta$ to $p_2=z_1^2+z_2^2+\dots:$ 
$$\Delta(p_2)=2+2+\dots,$$ which does not make sense in infinite dimension.
The solution is to {\bf define} this to be $2p_0,$ where $p_0$ is an additional parameter.

As well as in the trigonometric case the rational Calogero-Moser operator (\ref{CMratio}) is defined uniquely as a differential operator polynomially depending on the parameter $p_0$ such that when $p_0=N$ is a natural number the following diagram is commutative:
$$
\begin{array}{ccc}
\Lambda&\stackrel{{\mathbf
L}_{k,p_0}^{(\infty)}}{\longrightarrow}&\Lambda\\ \downarrow
\lefteqn{\varphi_{N}}& &\downarrow \lefteqn{\varphi_{N}}\\
\Lambda_{N}&\stackrel{{\mathbf
L}_k^{(N)}}{\longrightarrow}&\Lambda_{N} \\
\end{array}
$$
The main difference from the trigonometric case is that this operator has the degree $-2$ in the sense that it decreases the degree of the symmetric functions by 2, while the operator (\ref{inf1}) preserves the degrees (so has the degree 0). This means that in the rational case we do not have a good spectral theory, which in trigonometric case leads to a very important notion of the Jack symmetric functions.
Note that the same duality (\ref{sym}) holds  for the operator (\ref{inf1ratio}) as well.

It is interesting that there exists an infinite-dimensional version of CM operator, which has the degree $-1.$ It appeared in \cite{SV3} as a part of the $BC_{\infty}$ Calogero-Moser operator.
Consider the rational (gauged) $BC_N$ Calogero-Moser operator 
\begin{equation}
\label{bc}
B_{k,l}^{(N)}= \Delta-2k\sum_{i<j}^N\left(\frac{\partial_{i}-\partial_{j}}{x_{i}-x_{j}}+\frac{\partial_{i}+\partial_{j}}{x_{i}+x_{j}}\right)-2l\sum_{i=1}^N\frac{\partial_{i}}{x_{i}},\, \partial_i = \frac{\partial}{\partial x_i}
\end{equation}
and rewrite it in the coordinates $z_i=x_i^2$ (dividing for convenience by 4):
\begin{equation}
\label{CMS12}
\mathcal{B}_{k,l}^{(N)}=\sum_{i=1}^N z_i \partial_i^2 -2k\sum_{i<j}^N\frac{z_i\partial_{i}-z_j\partial_{j}}{z_{i}-z_{j}}-(l-1/2)\sum_{i=1}^N \partial_i, \, \partial_i = \frac{\partial}{\partial z_i}.
  \end{equation}
One can check that the corresponding infinite-dimensional analogue can be given in the same notations as above by 
$${ \mathcal B}_{k,l,p_0}^{(\infty)}=\sum_{a,b\geq 1}p_{a+b-1}\partial_{a}\partial_{b}-k\sum_{a,b\geq 1}p_{a}p_b \partial_{a+b+1} +(1+k)\sum_{a \geq 1}a p_{a-1}\partial_a $$
 \begin{equation}\label{inf2}
- (2k p_0+l+1/2)\sum_{a \geq 1}p_{a-1}\partial_a + kp_0^2 \partial_1
\end{equation}
(see formula (32) from \cite{SV3}). The duality for this operator has the form
\begin{equation}\label{sym6}
\theta^{-1} \circ \mathcal B_{k,l,p_0}^{(\infty)} \circ \theta = k \mathcal B_{ \bar k,\bar l, \bar p_0}^{(\infty)},
 \end{equation}
where $\bar k=k^{-1}, \bar p_0=k^{-1}p_0$ and $\bar l$ is defined by the relation
$$(2\bar l+1)=k^{-1}(2l+1).$$

\section{CMS operators in infinite dimension: Laurent version}

The relation of the theory of CMS operators with Lie superalgebras \cite{SV2} suggests the following Laurent extension of the CMS operators. Note that the finite dimensional CMS operators (\ref{CM}), (\ref{CMratio}) preserves the algebra of symmetric Laurent polynomials 
$$\Lambda^{\pm}_{N}=\Bbb C[z_{1}^{\pm1},\dots, z_{N}^{\pm1}]^{S_N}.$$
It is easy to see that it is generated (not freely) by power sums $p_{\pm 1}, \dots, p_{\pm N},$
where $$p_j(z) = z_1^j + \dots + z_N^j, \, j \in \mathbb Z.$$

Let us define its infinite-dimensional version $\Lambda^{\pm}$ as the commutative algebra with the free generators $p_{i},\, i\in \mathbb Z.$ It has a natural $\mathbb Z$-grading, where the degree of $p_i$ is $i.$ There is an involution $*$ such that $$p_i^* = p_{-i}.$$ This algebra can be also represented as
$ \Lambda^{\pm} = \Lambda^+ \otimes \Lambda^{-}\otimes \mathbb C[p_0],$
where $\Lambda^+$ is generated by $p_i$ with positive $i$ and $\Lambda^-$ by $p_i$ with negative $i.$ Note that the involution $*$ swaps $\Lambda^+$ and $\Lambda^-$ leaving $\mathbb C[p_0]$ fixed.

For every natural $N$ there is a homomorphism $\varphi_N:  \Lambda^{\pm} \rightarrow \Lambda^{\pm}_{N}:$
$$\varphi_N (p_i)= z_1^j + \dots + z_N^j, \, j \in \mathbb Z.$$
In particular, $\varphi_{N}(p_{0})=N$ is the dimension. The involution $*$ under this homomorphism goes to the natural involution on $\Lambda^{\pm}_{N}$ mapping $z_i$ to $z_i^{-1}.$

 \begin{thm}\label{main1} 
 {\em i)} There exists  a unique second order differential operator $\mathcal L_{k, p_0}^{(\pm \infty)}: \Lambda^{\pm} \to \Lambda^{\pm}$ polynomially dependent on $p_0$, such that for all $N=1,2,\dots$ and $p_0=N$   the following diagram is commutative
$$
\begin{array}{ccc}
\Lambda^{\pm}&\stackrel{{\mathcal
L}_{k,p_0}^{(\pm \infty)}}{\longrightarrow}&\Lambda^{\pm}\\ \downarrow
\lefteqn{\varphi_{N}}& &\downarrow \lefteqn{\varphi_{N}}\\
\Lambda_{N}^{\pm}&\stackrel{{\mathcal
L}_k^{(N)}}{\longrightarrow}&\Lambda_{N}^{\pm} \\
\end{array}
$$
where  $\mathcal L_k^{(N)}$ is the CMS operator (\ref{CM}).

 {\em ii)} Operator $\mathcal L_{k,p_0}^{(\pm\infty)}$ has the following explicit form:
$$
{ \mathcal L}_{k,p_0}^{(\pm \infty)}=\sum_{a,b\in \mathbb Z}p_{a+b}\partial_{a}\partial_{b}-k[\sum_{a,b>0}p_{a}p_b \partial_{a+b} -\sum_{a,b<0}p_{a}p_b \partial_{a+b}]
$$
 \begin{equation}\label{inf6}
 - kp_0 [\sum_{a>0} p_{a} \partial_{a}-\sum_{a<0} p_{a} \partial_{a}] +(1+k)\sum_{a\in\mathbb Z}a p_a\partial_a,
\end{equation}
where as before $\partial_a = a\partial/\partial p_a.$

{\em iii)} Operator $\mathcal L_{k,p_0}^{(\pm\infty)}$ has the following symmetries:
 \begin{equation}\label{sym12}
\theta^{-1} \circ \mathcal L_{k,p_0}^{(\pm\infty)} \circ \theta = k \mathcal L_{k^{-1}, k^{-1}p_0}^{(\pm\infty)},
 \end{equation}
where $\theta: p_a \rightarrow k  p_a$, and
\begin{equation}\label{sym13}
\mathcal L_{k,p_0}^{(\pm\infty) *} = \mathcal L_{k,p_0}^{(\pm\infty)}
\end{equation}
with respect to $*$-involution.
 \end{thm}

The proof is similar to the previous case. Note that the operator (\ref{inf6}) is well-defined on $\Lambda^{\pm}$ and that the parameter $p_0$ can not be eliminated by adding a suitable quantum integral (in contrast to the polynomial case).

\section{Jack--Laurent symmetric functions}

The Laurent polynomial eigenfunctions for CMS operators were first considered already by Sutherland in \cite{Suth}. They were later discussed in more details by Sogo \cite{Sogo1, Sogo2, Sogo3}, who  parametrized these eigenfunctions by the so-called extended Young diagrams, when the negative entries are also allowed. Alternatively, one can use two Young diagrams, corresponding to positive and negative parts.  However, in finite dimension this always can be reduced to the usual Jack polynomials simply by multiplication by a suitable power of the determinant $\Delta=z_1\dots z_N$
(see e.g. Forrester's comment in his MathSciNet review of the paper \cite{Sogo1}).

In infinite dimension this is however not possible because the product of $z_i$ does not belong to our algebra. This should be linked with Sogo's observation \cite{Sogo2} that the Laurent eigenfunctions are not stable and substantially depend on the dimension.

We define the {\it Jack--Laurent symmetric functions} $P_{\lambda,\mu} \in \Lambda^{\pm}$ labelled by two Young diagrams $\lambda$ and $\mu$ as follows.

Introduce the following  partial {\it dominance order} on the pairs of partitions $\lambda=(\lambda_1, \lambda_2, \dots, \lambda_l)$ and $\mu=(\mu_1, \mu_2, \dots, \mu_k).$ The weight of the pair $(\lambda,\mu)$ is defined as
$$|\lambda|-|\mu| = \sum_{i=1}^L \lambda_i -\sum_{j=1}^k \mu_j.$$ Another pair $(\tilde\lambda, \tilde\mu)$ with the same weight
$$|\lambda|-|\mu| = |\tilde\lambda|-|\tilde\mu|$$
satisfies the order relation
$$(\tilde\lambda, \tilde\mu) \preccurlyeq (\lambda, \mu)$$
if for all $n=1,2,...$
$$\tilde \lambda_1 +\dots +\tilde\lambda_n \leq \lambda_1+\dots+\lambda_n$$
and the same for $\mu.$
For given $\lambda$ and $\mu$ define 
$$\chi=(\lambda_{1},\dots,\lambda_{l},0,\dots,0,-\mu_{k},\dots,-\mu_{1}) \in \mathbb Z^N$$
and consider the corresponding Laurent  monomial symmetric polynomial 
 $$
 m_{\chi}=\sum z_{1}^{a_1}\dots z_{N}^{a_{N}},
 $$
 where sum is taken over all distinct permutations of $\chi=(\chi_1,\dots, \chi_N).$
 One can show that for any two partitions  such that $l(\lambda)+\l(\mu)\le N$ there exist a unique element $m_{\lambda,\mu}\in\Lambda^{\pm}$ such that
 $$
 \varphi_N(m_{\lambda,\mu})=m_{\chi}(z_{1},\dots, z_{N}).
 $$
 For example, in the simplest case $\lambda=(\lambda_1)$ and $\mu=(\mu_1)$
 $$ m_{\lambda,\mu}=p_{\lambda_1}p_{-\mu_1}-p_{\lambda_1-\mu_1}= p_{\lambda_1}p^*_{\mu_1}-p_{\lambda_1-\mu_1}.$$ In particular, when $\lambda_1=\mu_1$ we have
  $$ m_{\lambda,\lambda}=p_{\lambda_1}p^*_{\lambda_1}-p_0,$$
  so in general the monomial functions may explicitly depend on $p_0$.
  One more example: for $\lambda=(1,1)=(1^2),\, \mu=(1)$ one can check that
  $$m_{1^2,1}= \frac{1}{2}(p_1^2-p_2)p_{-1}-(p_0-1)p_1.$$

The monomial functions $m_{\lambda,\mu}$ give a convenient linear basis in $\Lambda^{\pm}.$
 The Jack--Laurent symmetric functions can be defined now in the usual way (cf. e.g. \cite{SV1}).

\begin{thm}
 {\em i)} If $k, p_0$ are generic then for  any pair of partitions $(\lambda,\mu)$  
 there exists a unique Laurent symmetric function
$P_{\lambda,\mu} =P_{\lambda,\mu}(k,p_0) \in \Lambda^{\pm}$ of the form
 \begin{equation}\label{tdef}
P_{\lambda,\mu}= m_{\lambda,\mu}+\sum_{(\lambda,\tilde\mu) \preccurlyeq (\lambda, \mu)}
u^{\tilde\lambda\tilde\mu}_{\lambda\mu}m_{\tilde\lambda,\tilde\mu}
\end{equation}
with some coefficients $u^{\tilde\lambda\tilde\mu}_{\lambda\mu}$, which is an eigenfunction of CMS
operator $\mathcal L_{k,p_0}^{(\pm\infty)}.$

 {\em ii)} The image $\varphi_N P_{\lambda,\mu}(k, N)$ for any $N>r+s, \, r=l(\lambda), s=l(\mu)$ can be expressed in terms of the usual Jack polynomials $P_{\nu}(z_1,\dots,z_N; k)$:
 \begin{equation}\label{jack}
 \varphi_N P_{\lambda,\mu}(k, N)= (z_1\dots z_N)^{-a} P_{\nu}(z_1,\dots,z_N; k),
\end{equation}
where $\nu=(\lambda_1+a, \dots, \lambda_{r}+a, a,\dots, a, a-\mu_s, \dots, a-\mu_1)$
for any $a \geq \mu_1.$ 
 
{\em iii)}  
Jack--Laurent symmetric functions obey the following $*$-duality property:
 \begin{equation}\label{du}
P_{\lambda,\mu}^*=P_{\mu,\lambda}.
\end{equation}
\end{thm}

To prove this we use the fact  that the CMS operator $\mathcal L_{k,p_0}^{(\pm\infty)}$ is triangular in 
the monomial basis $m_{\lambda,\mu}$ with respect to the dominance order defined above. The diagonal elements have the form
 \begin{equation}\label{dia}
E_{\lambda,\mu}= \sum \lambda_i^2 + \sum \mu_j^2 + k\sum (2i-1)\lambda_i + k\sum (2j-1)\mu_j - kp_0(|\lambda|+|\mu|).
\end{equation}
If $E_{\lambda,\mu}=E_{\tilde\lambda,\tilde\mu}$ for generic $k$ and $p_0$ then
$|\lambda|+|\mu|=|\tilde\lambda|+|\tilde\mu|.$ Since by definition
$|\lambda|-|\mu|=|\tilde\lambda|-|\tilde\mu|$ this implies that
$|\lambda|=|\tilde\lambda|$ and $|\mu|=|\tilde\mu|.$  The rest of the arguments are the same as in the usual case (see Section 2 in \cite{SV1}, page 346).

Formula (\ref{jack}) follows from the first part of Theorem 3.1 and shows that in finite dimensions everything can be reduced to the usual Jack polynomials.

For the eigenvalues of the suitable higher quantum CMS integrals $\mathcal L_s,\, \mathcal L_2 = \mathcal L_{k,p_0}^{(\pm\infty)}$ there exists a simple formula as a sum over boxes in the corresponding Young diagrams.

\begin{thm}
One can choose the quantum integrals $\mathcal L_s$ in such a way that 
$$\mathcal L_s P_{\lambda,\mu}=E^{(s)}_{\lambda,\mu} P_{\lambda,\mu}$$
with the eigenvalues $E^{(s)}_{\lambda,\mu}$ given by the formula
 \begin{equation}\label{eigen}
\sum_{(i,j)\in \lambda} \left[(j-\frac{1}{2})+k(i-\frac{1}{2})-\frac{kp_0}{2}\right]^{s-1}
+(-1)^s \sum_{(i,j)\in \mu} \left[(j-\frac{1}{2})+k(i-\frac{1}{2})-\frac{kp_0}{2}\right]^{s-1}.
\end{equation}
\end{thm}

The integrals $\mathcal L_s$ are non-stable versions of the quantum CMS integrals, corresponding to the Bernoulli sums (see formula (9) in \cite{SV1}).

Here is the explicit form of the Jack--Laurent symmetric functions in the simplest case:
 \begin{equation}\label{p11}
P_{1,1}(k,p_0)= p_1 p_{-1} - \frac{p_0}{1+k-kp_0},
\end{equation}
 \begin{equation}\label{p111}
P_{1^2,1}(k,p_0)=\frac{1}{2}(p_1^2-p_2)p_{-1} - \frac{2(p_0-1)}{2+4k-2kp_0}p_1,
\end{equation}
where $1^2$ denotes partition $\lambda=(1,1).$ In general, to compute  $P_{\lambda,\mu}$ one can use their definition, but the explicit form of the matrix elements $u^{\tilde\lambda\tilde\mu}_{\lambda\mu}$ is still to be found (cf. \cite{Sogo2}, section IV).

\section{Pieri formula for Jack--Laurent symmetric functions}

Let $\lambda$ and $\mu$ are two diagrams. 
Define for any positive integers $i,j$ the following functions
$$
c_{\lambda}(ji,x)=\lambda_{i}-j-k(\lambda^{\prime}_{j}-i)+x,\;\;
$$
$$
c_{\lambda\mu}(ji,x)=\lambda_{i}+j+k(\mu^{\prime}_{j}+i)+x.\;\;
$$
Consider all partitions $\tilde\lambda,$ which can be obtained by adding one box to $\lambda$, and define
\begin{equation}\label{V1}
V(\tilde\lambda\mid\lambda,\mu)=\prod_{r=1}^{i-1}\frac{c_{\lambda}(jr,1)c_{\lambda}(jr,-2k)}{c_{\lambda}(jr,-k)c_{\lambda}(jr,1-k)},
\end{equation}
where $(ij)$ is the added box.
Similarly, let $\tilde\mu$ be any partition, which can be obtained by deleting one box $(ij)$ from $\mu$ and define
$$
V(\tilde\mu\mid\lambda,\mu)=\prod_{r=i+1}^{l(\mu)}\frac{c_{\mu}(jr,1+k)c_{\mu}(jr,-k)}{c_{\mu}(jr,1)c_{\mu}(jr,0)}
$$
$$\times\prod_{r=1}^{l(\lambda)}\frac{c_{\lambda\mu}(jr,-1-k(p_{0}+2))c_{\lambda\mu}(jr,-kp_{0})}{c_{\lambda\mu}(jr,-1-k(p_{0}+1))c_{\lambda\mu}(jr,-k(p_{0}+1))}
$$
\begin{equation}\label{V2}
\times\frac{(j-1+k(l(\lambda)+\mu^{\prime}_{j}-p_{0}-1)(j+k(\mu^{\prime}_{j}-l(\mu)))}
{(j+k(l(\lambda)+\mu^{\prime}_{j}-p_{0}))(j-1+k(\mu^{\prime}_{j}-l(\mu)-1))},
\end{equation}
where $l(\lambda)$ is the {\it length}, which is the number of non-zero parts in the partition $\lambda.$

\begin{thm}
The Jack--Laurent symmetric functions satisfy the following Pieri formula:
\begin{equation}\label{pieri1}
p_{1}P_{\lambda,\mu}=\sum_{\tilde\lambda\supset\lambda}V(\tilde\lambda\mid\lambda,\mu)P_{\tilde\lambda,\mu}+\sum_{\tilde\mu\subset\mu}V(\tilde\mu,\lambda,\mu)P_{\lambda,\tilde\mu},
\end{equation}
where the sum is over all corresponding partitions $\tilde\lambda$, $\tilde\mu$ and the coefficients $V(\tilde\lambda\mid\lambda,\mu), \, V(\tilde\mu\mid\lambda,\mu)$ given by (\ref{V1}), (\ref{V2}).
\end{thm}

The proof follows from the usual Pieri formula for Jack polynomials \cite{Ma}.

One can rewrite the formula in terms of the following diagramatic representation of a pair partitions
(cf. \cite{MVDJ}).
For two partitions $\lambda,\mu$ consider the following geometric figure $Y=Y_{\lambda,\mu}=Y_{\lambda}\cup Y_{-\mu}\cup \Pi_{\lambda,\mu},$
where
$$
Y_{\lambda}=\{(ji)\mid\; j,i\in\Bbb Z,\;1\le i\le l(\lambda),\, 1\le j\le \lambda_{i}\},
$$
$$
Y_{-\mu}=\{(ji)\mid\; j,i\in\Bbb Z,\;-l(\mu)\le i\le -1,\, -\mu_{i}\le j\le -1\}
$$
and
$$
\Pi_{\lambda,\mu}=\{(ji)\mid\; j,i\in\Bbb Z,\;1\le i\le l(\lambda),\, - l(\mu^{\prime})\le j\le -1\}.
$$
On Fig. 1 we have the corresponding representation of $\lambda=(6,5,4,2,1)$ and $\mu=(7,3,2,1,1).$
Note that for $\lambda$ we follow the French way of drawing Young diagram, for $\mu$ it is rotated by 180 degrees.

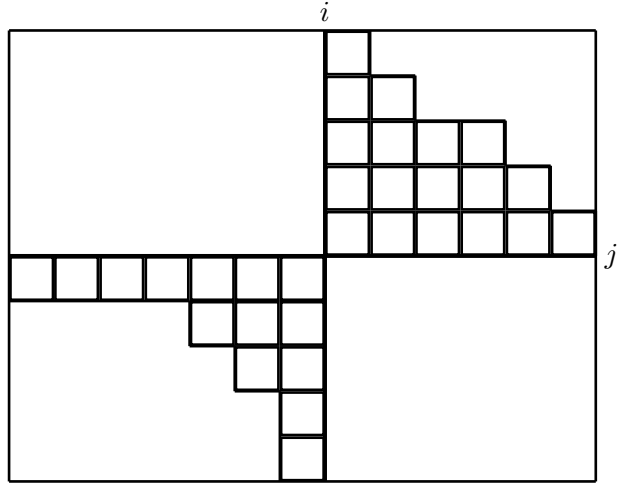
\begin{figure}[htbp]
\setlength{\unitlength}{0.06em}
\begin{center}
$$
{\psset{unit=0.3}
\begin{pspicture}(8,-8)(-8,10)
\newcommand{\OXe}{%
\psframe[linewidth=1pt, fillstyle=solid,fillcolor=white](-1,-1)(1,1) }
\newcommand{\OXsl}{%
\psframe[linewidth=1pt, fillstyle=crosshatch](-1,-1)(1,1) }
\newcommand{\OXgrey}{%
\psframe[linewidth=1pt, fillstyle=solid,fillcolor=green](-1,-1)(1,1)}
\newcommand{\OXbl}{%
\psframe[linewidth=1pt, fillstyle=solid,fillcolor=black](-1,-1)(1,1) }
%\pcline[linewidth=0.5pt](-1,-1)(1,1)
%\pcline[linewidth=0.5pt](1,-1)(-1,1)

\rput(1,1){\OXe} 
\rput(1,3){\OXe} 
\rput(1,5){\OXe} 
\rput(1,7){\OXe} 
\rput(1,9){\OXe} 
\rput(3,1){\OXe} 
\rput(3,3){\OXe} 
\rput(3,5){\OXe} 
\rput(3,7){\OXe} 
\rput(5,1){\OXe}
\rput(5,3){\OXe}
%\rput(5,3){\makebox(0,0){$\pi_1$}} 	
\rput(5,5){\OXe}	
%\rput(1,3){\makebox(0,0){$\pi_1$}} 
%\rput(5,7){\OXsl}	
\rput(7,1){\OXe} 
\rput(7,3){\OXe} 
\rput(7,5){\OXe} 
\rput(9,1){\OXe} 
\rput(9,3){\OXe} 
\rput(11,1){\OXe}

\rput(-1,-1){\OXe} 
\rput(-1,-3){\OXe} 
\rput(-1,-5){\OXe} 
\rput(-1,-7){\OXe} 
\rput(-1,-9){\OXe}
\rput(-3,-1){\OXe} 
\rput(-3,-3){\OXe} 
\rput(-3,-5){\OXe} 
\rput(-5,-1){\OXe} 
\rput(-5,-3){\OXe} 

\rput(-7,-1){\OXe} 
\rput(-9,-1){\OXe} 
\rput(-11,-1){\OXe} 
\rput(-13,-1){\OXe} 

\pcline[linewidth=1pt](12,0)(12,2)
\pcline[linewidth=1pt](12,2)(10,2)
\pcline[linewidth=1pt](10,2)(10,4)
\pcline[linewidth=1pt](10,4)(8,4)
\pcline[linewidth=1pt](8,4)(8,6)
\pcline[linewidth=1pt](8,6)(4,6)
\pcline[linewidth=1pt](4,6)(4,8)
\pcline[linewidth=1pt](4,8)(2,8)
\pcline[linewidth=1pt](2,8)(2,10)

\pcline[linewidth=1pt](-14,-2)(-6,-2)
\pcline[linewidth=1pt](-6,-2)(-6,-4)
\pcline[linewidth=1pt](-6,-4)(-4,-4)
\pcline[linewidth=1pt](-4,-4)(-4,-6)
\pcline[linewidth=1pt](-4,-6)(-2,-6)
\pcline[linewidth=1pt](-2,-6)(-2,-10)
 
%\rput(9,5){\OXe} 
%\rput(3,1){\OXe}
%\rput(3,3){\OXsl} 
\pcline[linewidth=1pt](-14,10)(12,10)
\pcline[linewidth=1pt](-14,10)(-14,-10)
\pcline[linewidth=1pt](-14,-10)(12,-10)
\pcline[linewidth=1pt](12,-10)(12,10)
\pcline[linewidth=1.5pt](-14,0)(12,0)
\pcline[linewidth=1.5pt](0,-10)(0,10)
\rput(0,10.4){\makebox(0,0)[cb]{$i$}}
\rput(12.4,0){\makebox(0,0)[lc]{$j$}}
\end{pspicture}
}
$$
\medskip
\caption{Diagrammatic representation of a pair of partitions}
\label{Fig:Diagr}
\end{center}
\end{figure}

Define the following analogues of rows
$$
y_{i}=\begin{cases}\lambda_{i},& 1\le i\le l(\lambda)\cr -\mu_{-i},& -l(\mu)\le i \le -1 \end{cases}
$$
and columns
$$
y^{\prime}_{j}=\begin{cases}\lambda^{\prime}_{j},& 1\le j\le l(\lambda^{\prime})\cr -\mu^{\prime}_{-j},& -l(\mu^{\prime})\le j \le -1 \end{cases}
$$
with all other $y_{i}$, $y^{\prime}_{j}$ being zero.
For every box $\Box$ with integer coordinates $(j,i)$ define the function 
$$
c_{Y}(\Box,x)=y_{i}-j-k(y^{\prime}_{j}-i)+x.
$$
Define for the added box $\Box=(j,i)$ the following subset in $Y_{\lambda}$
$$
\pi_{1}=\{(j,r)\mid 1\le r< i\}
$$ 
and  for  deleted  box  $\Box=(ji)$  the subsets in $Y$
$$
\pi_{2}=\{(j,r)\mid -l(\mu)  \le r<-\mu^{\prime}_{-j} \},
$$
$$
\pi_{3}=\{(j,r)\mid 1 \le r\le l(\lambda)\}.
$$
The meaning of these subsets is clear from Fig. 2, where the deleted box is black and the added box is crosshatched.

\begin{figure}[htbp]
\setlength{\unitlength}{0.06em}
\begin{center}
$$
{\psset{unit=0.3}
\begin{pspicture}(8,-8)(-8,10)
\newcommand{\OXe}{%
\psframe[linewidth=1pt, fillstyle=solid,fillcolor=white](-1,-1)(1,1) }
\newcommand{\OXsl}{%
\psframe[linewidth=1pt, fillstyle=crosshatch](-1,-1)(1,1) }
\newcommand{\OXgrey}{%
\psframe[linewidth=1pt, fillstyle=solid,fillcolor=green](-1,-1)(1,1)}
\newcommand{\OXbl}{%
\psframe[linewidth=1pt, fillstyle=solid,fillcolor=black](-1,-1)(1,1) }
%\pcline[linewidth=0.5pt](-1,-1)(1,1)
%\pcline[linewidth=0.5pt](1,-1)(-1,1)

\rput(1,1){\OXe} 
\rput(1,3){\OXe} 
\rput(1,5){\OXe} 
\rput(1,7){\OXe} 
\rput(1,9){\OXe} 
\rput(3,1){\OXe} 
\rput(3,3){\OXe} 
\rput(3,5){\OXe} 
\rput(3,7){\OXe} 
\rput(5,1){\OXgrey}
\rput(5,3){\OXgrey}
\rput(5,3){\makebox(0,0){$\pi_1$}} 	
\rput(5,5){\OXgrey}	
%\rput(1,3){\makebox(0,0){$\pi_1$}} 
\rput(5,7){\OXsl}	
\rput(7,1){\OXe} 
\rput(7,3){\OXe} 
\rput(7,5){\OXe} 
\rput(9,1){\OXe} 
\rput(9,3){\OXe} 
\rput(11,1){\OXe} 

\rput(-1,-1){\OXe} 
\rput(-1,-3){\OXe} 
\rput(-1,-5){\OXe} 
\rput(-1,-7){\OXe} 
\rput(-1,-9){\OXe}
\rput(-3,-1){\OXe} 
\rput(-3,-3){\OXe} 
\rput(-3,-5){\OXe} 
\rput(-5,-1){\OXe} 
\rput(-5,-3){\OXbl} 
\rput(-5,-5){\OXgrey}
\rput(-5,-7){\OXgrey}
\rput(-5,-7){\makebox(0,0){$\pi_2$}} 	
\rput(-5,-9){\OXgrey}
\rput(-7,-1){\OXe} 
\rput(-9,-1){\OXe} 
\rput(-11,-1){\OXe} 
\rput(-13,-1){\OXe} 

\rput(-5,1){\OXgrey}
\rput(-5,3){\OXgrey}
\rput(-5,5){\OXgrey}
\rput(-5,5){\makebox(0,0){$\pi_3$}} 	
\rput(-5,7){\OXgrey}
\rput(-5,9){\OXgrey}

\pcline[linewidth=1pt](12,0)(12,2)
\pcline[linewidth=1pt](12,2)(10,2)
\pcline[linewidth=1pt](10,2)(10,4)
\pcline[linewidth=1pt](10,4)(8,4)
\pcline[linewidth=1pt](8,4)(8,6)
\pcline[linewidth=1pt](8,6)(4,6)
\pcline[linewidth=1pt](4,6)(4,8)
\pcline[linewidth=1pt](4,8)(2,8)
\pcline[linewidth=1pt](2,8)(2,10)

\pcline[linewidth=1pt](-14,-2)(-6,-2)
\pcline[linewidth=1pt](-6,-2)(-6,-4)
\pcline[linewidth=1pt](-6,-4)(-4,-4)
\pcline[linewidth=1pt](-4,-4)(-4,-6)
\pcline[linewidth=1pt](-4,-6)(-2,-6)
\pcline[linewidth=1pt](-2,-6)(-2,-10)

%\rput(9,5){\OXe} 
%\rput(3,1){\OXe}
%\rput(3,3){\OXsl} 
\pcline[linewidth=1pt](-14,10)(12,10)
\pcline[linewidth=1pt](-14,10)(-14,-10)
\pcline[linewidth=1pt](-14,-10)(12,-10)
\pcline[linewidth=1pt](12,-10)(12,10)
\pcline[linewidth=1.5pt](-14,0)(12,0)
\pcline[linewidth=1.5pt](0,-10)(0,10)
\rput(0,10.4){\makebox(0,0)[cb]{$i$}}
\rput(12.4,0){\makebox(0,0)[lc]{$j$}}
\end{pspicture}
}
$$
\medskip
\caption{Summation sets for the Pieri formula}
\label{Fig:Diagr2}
\end{center}
\end{figure}
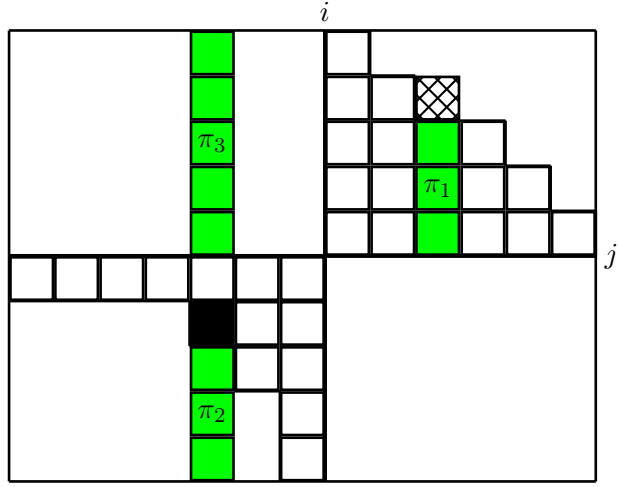

In these terms the coefficients in the Pieri formula (\ref{pieri1}) can be written as

\begin{equation}\label{V11}
V(\tilde\lambda,\lambda,\mu)=  \prod_{\Box\in\pi_1}\frac{c_{Y}(\Box,-2k)c_{Y}(\Box,1)}{c_{Y}(\Box,-k)c_{Y}(\Box,1-k)},
\end{equation}
$$
V(\tilde\mu,\lambda,\mu)=  \prod_{\Box\in\pi_2}\frac{c_{Y}(\Box,-1-k)c_{Y}(\Box,k)}{c_{Y}(\Box,-1)c_{Y}(\Box,0)}
$$
$$
\times\prod_{\Box\in\pi_3}\frac{c_{Y}(\Box,-1-k(p_{0}+2))c_{Y}(\Box,-kp_{0})}{c_{Y}(\Box,-1-k(p_{0}+1))c_{Y}(\Box,-k(p_{0}+1))}
$$
\begin{equation}\label{V21}
\times\frac{(j+1+k(y^{\prime}_{j}-y^{\prime}_{1}-p_{0}-1)(j+k(y^{\prime}_{j}+y^{\prime}_{-1})}
{(j+k(y^{\prime}_{j}-y^{\prime}_{1}-p_{0}))(j+1+k(y^{\prime}_{j}+y^{\prime}_{-1}+1)}
\end{equation}
with the convention that the product over empty set is equal to 1.

A non-symmetry between $\lambda$ and $\mu$ is due to the choice of $p_1$ in the left hand side of the Pieri formula. By applying $*$-involution to formula (\ref{pieri1}) one has the corresponding formula for $p_{-1}$, where the roles of $\lambda$ and $\mu$ are interchanged. 

Another remark is that in the Pieri formula one can replace the rectangle containing figure $Y$ by any bigger rectangle with the simultaneous change of $y_1^{\prime}$ and $y_{-1}^{\prime}$ by the corresponding lengths in positive and negative directions.

\section{One more extension}

The following extension is motivated by the relations with Grothendieck ring of Lie superalgebra $\mathfrak{gl}(m,n)$ (see \cite{SV3}). In this ring we have the element $\Delta,$ corresponding to the Berezinian, which does not belong to the algebra generated by the (super) power sums (in contrast to the usual case of $\mathfrak{gl}(m)$). It can be considered as the image of the following new formal variable, which we denote $w.$

Consider the polynomial extension $\Lambda^{\pm}[w]$ of the Laurent algebra $\Lambda^{\pm}$ considered above. Introduce the following extension of the CMS operator
$$
\hat {\mathcal L}_{k,p_0}^{(\pm \infty)}=\sum_{a,b\in \mathbb Z}p_{a+b}\partial_{a}\partial_{b}-k[\sum_{a,b>0}p_{a}p_b \partial_{a+b} -\sum_{a,b<0}p_{a}p_b \partial_{a+b}]
$$
 \begin{equation}\label{inf12}
 - kp_0 [\sum_{a>0} p_{a} \partial_{a}-\sum_{a<0} p_{a} \partial_{a}] +(1+k)\sum_{a\in\mathbb Z}a p_a\partial_a +p_0(w \partial_w)^2 + 2w \partial_w \sum_{a\in \mathbb Z} p_a\partial_a,
\end{equation}
where  $\partial_a = a\frac{\partial}{\partial p_a}$ and $\partial_w = \frac{\partial}{\partial w}.$

Let us extend also the homomorphism $\varphi_N$ to 
$\hat\varphi_N: \Lambda^{\pm}[w] \rightarrow \Lambda_N$ 
by $$\hat\varphi_N(w)=z_1\dots z_N.$$

 \begin{thm}\label{main1} 
The following diagram is commutative
$$
\begin{array}{ccc}
\Lambda^{\pm}[w]&\stackrel{\hat{\mathcal
L}_{k,p_0}^{(\pm \infty)}}{\longrightarrow}&\Lambda^{\pm}[w]\\ \downarrow
\lefteqn{\hat\varphi_{N}}& &\downarrow \lefteqn{\hat\varphi_{N}}\\
\Lambda_{N}^{\pm}&\stackrel{{\mathcal
L}_k^{(N)}}{\longrightarrow}&\Lambda_{N}^{\pm}, \\
\end{array}
$$
where  $\mathcal L_k^{(N)}$ is the CMS operator (\ref{CM}).
\end{thm}

This operator preserves the double grading in both $p_a$ and $w$ (alternatively, one can define the degree of $w$ to be $p_0$). Its eigenfunctions
have the form
$$P_{\lambda,\mu,l}= w^l P_{\lambda,\mu},$$
where $l\in\mathbb Z_{\geq 0}$ and $P_{\lambda,\mu}$ are the Jack-Laurent symmetric functions:
$$\hat {\mathcal L}_{k,p_0}^{(\pm \infty)}P_{\lambda,\mu,l}= E_{\lambda,\mu,l} P_{\lambda,\mu,l},$$
where 
$$E_{\lambda,\mu,l}=E_{\lambda,\mu} +p_0 l^2 + 2l (|\lambda|-|\mu|)$$
with $E_{\lambda,\mu}$ given by (\ref{dia}). 

\section{Special case $k=-1$ and Lie superlagebra $\mathfrak{gl}(m,n)$}

The case when $k=-1$ is special as one can see already from the explicit form of the simplest Jack-Laurent symmetric function (\ref{p11}). However, similarly to $BC$-case \cite{SV4}, the limit of Jack-Laurent symmetric functions $P_{\lambda,\mu}(k,p_0)$ when $k\rightarrow -1$ is well defined for {\it generic} values of parameter $p_0.$ We claim that the corresponding limit $P_{\lambda,\mu}(-1)$
actually does not depend on $p_0$ as it follows from the following Jacobi--Trudy formula.

Let $h_i \in \Lambda \subset \Lambda^{\pm}, \,i \in \mathbb Z$ be the complete symmetric functions \cite{Ma} for $i \geq 0$ and $h_i =0$ for $i<0.$ Define 
$h_i^*$ as the image of $h_i$ under the $*$-involution in $\Lambda^{\pm}.$

 \begin{thm}\label{main123}
 The limit of Jack-Laurent symmetric functions $P_{\lambda,\mu}(k,p_0)$ when $k\rightarrow -1$ for generic $p_0$ does not depend on  $p_0$ and can be given by the following Jacobi--Trudy formula
 as $(r+s)\times(r+s)$ determinant, where $r=l(\lambda),\, s=l(\mu)$ are the number of parts in $\lambda$ and $\mu$:
\begin{equation}
\label{JT}
P_{\lambda,\mu}(-1)=\left|\begin{array}{cccc}
   h^*_{\mu_{s}}&h^{*}_{\mu_{s-1}}& \ldots &h^{*}_{\mu_{s}-s-r+1}\\
\vdots&\vdots&\ddots&\vdots\\
  h^*_{\mu_{1}+s-1}&h^{*}_{\mu_{1}+s-2}& \ldots &h^{*}_{\mu_{1}-r}\\
  h_{\lambda_{1}-s}&h_{\lambda_{1}-s+1}& \ldots &h_{\lambda_{1}+r-1}\\
\vdots&\vdots&\ddots&\vdots\\
  h_{\lambda_{r}-s+r-1}&h_{\lambda_{r}-s+r}& \ldots &h_{\lambda_{r}}\\
 \end{array}\right|
 \end{equation}
\end{thm}

The proof follows from the formula (\ref{jack}) and the usual Jacobi--Trudy formula for Schur polynomials.

Consider the homomorphism $\varphi_{m,n}: \Lambda^{\pm} \rightarrow \Lambda_{m,n}^{\pm}$ defined by
$\phi_{m,n}(p_i)= p_i(x,y),$ where
$$
p_i(x,y)=x_1^i+\dots+x_m^i-y_1^i-\dots-y_n^i, \, i \in \mathbb Z
$$
are the corresponding super version of the power sums. In particular, $p_0$ is specialised to the superdimension $m-n.$ 

The algebra $\Lambda_{m,n}^{\pm}$ is defined here as the subalgebra of Laurent polynomials in $x$ and $y$ generated by $p_i(x,y).$ It appears as a part of the Grothendieck ring of the Lie superalgebra $\mathfrak{gl}(m|n),$ which is generated by $p_i(x,y)$, $\Delta=\frac{x_1\dots x_m}{y_1\dots y_n}$ and $\Delta^*=\frac{y_1\dots y_n}{x_1\dots x_m}$ (see \cite{SV3}).

The image of the functions $P_{\lambda,\mu}(-1)$ under the homomorphism $\varphi_{m,n}$ coincide with the functions
$s_{\bar \mu, \lambda}(x/y)$ from the paper \cite{MVDJ} by Moens and Van der Jeugt. This follows from the comparison of (\ref{JT}) with the formula (2.3) in  \cite{MVDJ}). Moens and Van der Jeugt showed that under certain conditions on the Young diagrams $\lambda,\mu$ the functions $s_{\bar \mu, \lambda}(x/y)$ give the characters of the irreducible modules of $\mathfrak{gl}(m,n),$ but in general the representation-theoretic interpretation of $s_{\bar \mu, \lambda}(x/y)$ is not known.

We expect that Euler supercharacters will appear naturally in this relation, similarly to the orthosymplectic $BC$-case considered in \cite{SV4}. To see what is happening one can look at the simplest example of the Jack--Laurent symmetric function (\ref{p11}). If we simply put $k=-1$ into (\ref{p11}) we have
$$P_{1,1}(-1,p_0)= p_1 p_{-1} - \frac{p_0}{p_0},$$
which is equal to 
$P_{1,1}(-1)= p_1 p_{-1}-1$ if $p_0\neq 0$ (or $m\neq n$ after applying the homomorphism $\varphi_{m,n}$). Note that when $m\neq n$ the image of $p_1 p_{-1}-1$ under $\varphi_{m,n}$ is the supercharacter of the quotient module $V^*\otimes V/\mathbb C,$ where $V$ is the standard representation of $\mathfrak{gl}(m,n).$
If $m=n$ this module is not irreducible and the conditions of Moens and Van der Jeugt are not satisfied.
We believe that in that case one should look at the corresponding Euler supercharacter, which in this example turns out to be the image of the element from the extended algebra $\Lambda^{\pm}[w]$
$$E_{1,1}=p_1 p_{-1}-1+w,$$  where we assume that $\varphi_{m,n}(w)=\Delta=\frac{x_1\dots x_m}{y_1\dots y_n}$ is the Berezinian. We hope that the infinite-dimensional CMS operators could help to clarify this.

\section{Acknowledgements}

We are grateful to Vladimir Bazhanov and Gregory Korchemsky for stimulating discussions and to Leonid Chekhov for the help with preparation of figures.

This work has been partially supported by EPSRC (grant EP/E004008/1) and by the
European Union through the FP6 Marie Curie RTN ENIGMA (contract
number MRTN-CT-2004-5652).

\end{document}